\documentclass[conference,a4paper]{IEEEtran}
\usepackage{xcolor}
\usepackage{balance}

\ifCLASSINFOpdf
  \usepackage[pdftex]{graphicx}
\else
  \usepackage[dvips]{graphicx}
\fi
\usepackage{amsmath,amssymb}
\usepackage{physics}
\ifCLASSOPTIONcompsoc
 \usepackage[caption=false,font=normalsize,labelfont=sf,textfont=sf]{subfig}
\else
 \usepackage[caption=false,font=footnotesize]{subfig}
\fi
\usepackage{cite}

\usepackage[colorlinks]{hyperref}
\usepackage{xcolor}
\hypersetup{
colorlinks,
    linkcolor={red!50!black},
    citecolor={blue!50!black},
    urlcolor={blue!80!black}
}

\usepackage{mathtools}

\def\XXint#1#2#3{{\setbox0=\hbox{$#1{#2#3}{\int}$}
     \vcenter{\hbox{$#2#3$}}\kern-.5\wd0}}

\usepackage{bm}
\newcommand{\R}{\mathbb{R}}
\newcommand{\C}{\mathbb{C}}

\newcommand{\ie}{\textit{i.e.}\/, }
\newcommand{\eg}{\textit{e.g.}\/, }
\newcommand{\cf}{\textit{cf.}\/, }

\providecommand*{\mrm}[1]{\mathrm{#1}}
\providecommand*{\unit}[1]{\ensuremath{\mrm{\,#1}}}
\providecommand*{\eu}{\ensuremath{\mrm{e}}}
\providecommand*{\iu}{\ensuremath{\mrm{i}}}
\providecommand*{\ju}{\ensuremath{\mrm{j}}}

\renewcommand{\Re}{\ensuremath{\mrm{Re}}}	
\renewcommand{\Im}{\ensuremath{\mrm{Im}}}	

\newcommand{\toh}{\hat{\to}}

\begin{document}
%
\title{Time-domain constraints for Positive Real functions: \\ Applications to the dielectric response of a passive material}

\author{\IEEEauthorblockN{
Sven~Nordebo\IEEEauthorrefmark{1},   
Martin~Stumpf\IEEEauthorrefmark{2},   
}                                     
\IEEEauthorblockA{\IEEEauthorrefmark{1}
Department of Physics and Electrical Engineering, Linn\ae us University,   351 95 V\"{a}xj\"{o}, Sweden. E-mail: sven.nordebo@lnu.se} 
\IEEEauthorblockA{\IEEEauthorrefmark{2}
Lerch Laboratory of EM Research, Department of Radio Electronics, FEEC, Brno University of Technology, \\ Technick\'{a} 3082/12, 616 00 Brno, The Czech Republic. E-mail: martin.stumpf@centrum.cz \\
The Department of Computer Science, Electrical and Space Engineering, EISLAB,  \\ Luleå University of Technology, 971 87 Luleå, Sweden. 
E-mail: martin.stumpf@ltu.se} 
}



\maketitle

\begin{abstract}
This paper presents a systematic approach to derive physical bounds for Positive Real (PR) functions directly in the Time-Domain (TD).
The theory is based on Cauer's representation of an arbitrary PR function together with  
associated sum rules (moments of the measure) and exploits the unilateral Laplace transform to derive rigorous bounds on the TD response of a passive system. 
The existence of useful sum rules and related physical bounds relies heavily on an 
assumption about the PR function having a low- or high-frequency asymptotic expansion at least of odd order 1.
As a canonical example, we explore the time-domain dielectric step response of a passive material, either with or without a given pulse raise time.
As a particular numerical example, we consider here the electric susceptibility of gold (Au) 
which is commonly modeled by well established Drude or Brendel Bormann models.  An explicit physical bound 
on the early-time step response of the material is then given in terms of a quadratic function in time which is completely determined
by the plasma frequency of the metal. 
\end{abstract}

\vskip0.5\baselineskip

\section{Introduction}

It is well known that the Kramers-Kronig relations 
limit the dispersive behavior of a linear, time-invariant and causal system\cite{Nussenzveig1972,King2009,Haakestad+Skaar2005}. 
The additional assumption of passivity may furthermore imply additional physical limitations on what is possible to realize in a finite bandwidth.
More precisely, we refer here to immittance passivity, which by itself implies that the system is causal\cite{Zemanian1965}.
Classical examples are the bounds on broadband matching using lossless networks
that were derived by Fano\cite{Fano1950}. More recent examples are the physical bounds that have been obtained
concerning radar absorbers\cite{Rozanov2000}, high-impedance surfaces\cite{Gustafsson+Sjoberg2011}, passive metamaterials\cite{Gustafsson+Sjoberg2010a,Skaar+Seip2006}, 
broadband quasi-static cloaking\cite{Cassier+Milton2017}, scattering\cite{Sohl+etal2007a,Bernland+etal2011b}, antennas\cite{Gustafsson+etal2010a}, 
reflection coefficients\cite{Gustafsson2010b}, waveguides\cite{Vakili+etal2014} and periodic structures\cite{Gustafsson+etal2012c}, etc. 
A survey of recent examples and applications in electromagnetics is given in\cite{Nedic+etal2019}.

The immittance passive systems can be completely characterized by Positive Real (PR) functions 
(analytic functions mapping the right half-plane into itself), or equivalently, by the so called (symmetric) Herglotz functions 
(analytic functions mapping the upper half-plane into itself), also known as Nevanlinna or Herglotz-Nevanlinna functions, 
\cf \eg\cite{Zemanian1965,Kac+Krein1974,Akhiezer1965,Nussenzveig1972,Gesztesy+Tsekanovskii2000}. 
Provided that a PR function has some odd ordered low- and/or high-frequency asymptotic expansion at least of order 1,
a partial knowledge about the expansion coefficients can then sometimes be used to derive sum rules (integral identities) which may have
a useful physical interpretation.  Physical bounds can then been obtained by 
restricting an integral  to a finite frequency interval and hence bounding it from above by the corresponding sum rule (moments of a positive measure),
see \eg\cite{Bernland+etal2011b,Nedic+etal2019}.
This technique, which has its roots from the 50's\cite{Fano1950}, has been given a solid foundation with theory and applications
as mentioned in the references given above. However, there can still be many interesting applications to explore
and what is still missing is an investigation about the  physical limitations of a passive system that can be formulated directly in the time-domain.

A new approach to derive physical bounds in the Time-Domain (TD) has recently been given in\cite{Stumpf-Nordebo:2024a,Stumpf-Nordebo:2024b}. 
This technique takes as its starting point the low- and high-frequency asymptotic properties of a linear, time-invariant and casual system
and exploits its analytical properties and the calculus of residues to derive physical bounds directly in the TD. In particular, by exploring
various subclasses of linear systems and their asymptotic properties together with some adequately chosen unipolar input pulses,
it has been demonstrated how new {\em early-time} as well as {\em late-time} physical bounds on the system response
can be derived.  Sometimes these bounds can furthermore be combined by their common {\em corner time} to provide useful {\em all-time} bounds. 
The purpose of this paper is to systematically explore these ideas by assuming
that the linear system is {\em immittance passive} and hence can be characterized by a PR function.
The present approach is based on Cauer's representation\cite{Zemanian1965} of an arbitrary PR function and its associated sum rules\cite{Nedic+etal2019}.
The unilateral Laplace transform is employed to obtain a theory which is directly applicable in the TD.

Typical application areas for TD physical bounds is within Electromagnetic Compatibility (EMC), security for revealing signals, transient protection (lightning etc)
as well as with high speed electronic and photonic switching circuits, etc., see \eg \cite{Stumpf:2019lightning,stumpf:2017shielding,Lager:2012} for further references.
As a canonical example we choose here to investigate
TD physical bounds on the dielectric response of a passive material, 
including conduction, Debye, Lorentz, Drude and the Brendel Bormann models\cite{Brendel+Bormann1992,Rakic+etal1998,Nordebo+etal2019a}. 
As a particular numerical example we will consider here the electric susceptibility of gold (Au) and give an explicit bound on its early-time step response
in terms of its plasma frequency which can be derived from well established Drude\cite{Olmon+etal2012} and Brendel Bormann models\cite{Rakic+etal1998}.
To this end, it is noted that a metal (in particular a Drude material) behaves as a conductor at low frequencies (late times) and 
increasingly as a dielectric at higher frequencies (early times).

The rest of the paper is organized as follows.
In Appendices \ref{sect:PRbasics} and \ref{sect:PRsumrules} are given a brief survey of the most important properties of PR functions
which are needed here.  
A general description of the time-domain
bounds for PR functions which can be derived based on their associated sum rules is given in section \ref{sect:TDPR}.
The theory is then specialized to the step response of a passive dielectric material in section \ref{sect:StepDielectric} 
with the electric susceptibility of gold (Au) and its plasma frequency as the main numerical example.
A summary with conclusions are finally given in section \ref{sect:Summary}.

\section{Time-domain constraints for Positive Real functions}\label{sect:TDPR}
Basic properties and sum rules for Positive Real (PR) functions are given in Appendix \ref{sect:PRbasics}.
Several time-domain constraints for PR functions can be derived based on Cauer's representation \eqref{eq:Cauer}
together with the sum rules \eqref{eq:Herglotz_SR1PR} and \eqref{eq:Herglotz_SR2PR}. For the cases where this is possible the corresponding
bounds are rigorous due to the strict positivity of the generating measure $\beta$. The number of feasible formulations are however restricted 
by the structure of the asymptotic expansions in \eqref{eq:PR_assympt1} and \eqref{eq:PR_assympt2}. Note in particular
the requirement of having an {\em odd} expansion  within the range of feasible sum rules for $n=0, 2, 4,\ldots$. 
Note also that we may have different expansion orders $M$ associated with \eqref{eq:PR_assympt1} and \eqref{eq:PR_assympt2}.
Below, we will demonstrate the procedure by explicitly deriving a number of useful constraints that are associated with the minimum required asymptotic order $M=1$.
Higher order constraints can be similarly derived if the necessary a priori information is available. 

\subsection{Early-time bounds} 
We start by rewriting Cauer's representation \eqref{eq:Cauer} for an arbitrary PR function as
\begin{equation}\label{eq:Cauer2}
p(s) =b_1s+a_{-1}s^{-1}+\int_{\R\setminus\{0\}}\frac{s}{s^2+\xi^2}\mrm{d}\beta(\xi),
\end{equation}
where $s$ is the ordinary Laplace variable with $\Re\{s\}>0$, $b_1\geq 0$, $a_{-1}\geq 0$ and $\beta(\xi)$ is a positive Borel measure
with growth condition given by \eqref{eq:growthcond}.
The inverse Laplace transform is then given by the following distribution of slow growth
\begin{equation}\label{eq:poft}
p(t)=b_1\delta^{(1)}(t)+a_{-1}H(t)+H(t)\int_{\R\setminus\{0\}}\cos(\xi t)\mrm{d}\beta(\xi),
\end{equation}
where $H(t)$ is the Heaviside unit step function and $\delta^{(1)}(t)$ the first order derivative of the Dirac delta function  $\delta(t)$, see also \cite[Theorem 10.5-1]{Zemanian1965}.
For notational convenience we let the argument of a function $f(\cdot)$ decide whether we refer to the time-domain $f(t)$ or to the Laplace domain $f(s)$, etc.
It is furthermore noticed that $p(t)$ corresponds to the impulse response of an immittance passive system and is always a causal function, see also \cite[Chapt. 10.3]{Zemanian1965}.

Let us now furthermore assume that there exists an odd ordered high-frequency asymptotic expansion of order 1
\begin{equation}\label{eq:highfrequency}
p(s)=\left\{\begin{array}{l}
a_{-1}s^{-1}+o(s^{-1}), \quad \textrm{as } s\toh 0 \vspace{0.2cm} \\
b_1s+b_{-1}s^{-1}+o(s^{-1}), \quad \textrm{as } s\toh\infty,
\end{array}\right.
\end{equation}
according to the definitions made in \eqref{eq:PR_assympt1} and \eqref{eq:PR_assympt2} 
and where $o(\cdot)$ denotes the small ordo according to the definition made in \eg \cite[p.~4]{Olver1997}.
We have then the following sum rule \eqref{eq:Herglotz_SR2PR} for $n=0$
\begin{equation}\label{eq:sumrulebm1am1}
\int_{\R\setminus\{0\}}\mrm{d} \beta(\xi)=b_{-1}-a_{-1}.
\end{equation}
From the positivity of the measure it is concluded that $b_{-1}-a_{-1}\geq 0$ and where $b_{-1}=a_{-1}$ corresponds to the trivial case for which the positive measure $\beta$ is 
different from zero on $\R\setminus\{0\}$ only at a set of measure zero. 
From \eqref{eq:poft} follows then that
\begin{multline}\label{eq:poft2}
\pm\left(p(t)-b_1\delta^{(1)}(t)-a_{-1}H(t)\right) \\
=\pm H(t)\int_{\R\setminus\{0\}}\cos(\xi t)\mrm{d}\beta(\xi) \\
\leq H(t)\int_{\R\setminus\{0\}}\mrm{d}\beta(\xi)=(b_{-1}-a_{-1})H(t),
\end{multline}
and where the inequality should be understood in the distributional sense.

We can now derive a simple early-time bound for the response of any right-sided and unipolar input pulse shape $f(t)\geq 0$ for $t\geq 0$ directly from \eqref{eq:poft2} as
\begin{equation}\label{eq:alltimebounds}
\left| p(t)*f(t)-b_1\delta_tf(t)-a_{-1}\delta_t^{-1}f(t)\right| \leq (b_{-1}-a_{-1})\delta_t^{-1}f(t)
\end{equation}
where $*$ denotes the time-domain convolution, $\delta_t$ the time derivative and 
\begin{equation}
\delta_t^{-1}f(t)=H(t)*f(t)=\int_{0-}^t f(\tau)\mrm{d}\tau,
\end{equation}
see also \cite{Stumpf-Nordebo:2024a}.
Even though the bound in \eqref{eq:alltimebounds} is an all-time bound valid for all $t>0$ 
we refer to it as an early-time bound as it is generally most accurate asymptotically as $t\rightarrow 0+$.
It is noted that the inequality in \eqref{eq:poft2}  is preserved under the convolution as the pulse shape $f(t)$ is assumed to be non-negative on its region of support $[0,\infty)$.
It is furthermore assumed that $f(t)$ is generally a distribution of slow growth and that its unilateral Laplace transform $f(s)$ exists.
To this end, the operator $\delta_t^{-n}$ will denote the time-domain integrator of order $n$ corresponding to 
a multiplication with $s^{-n}$ in the Laplace-domain. 
It should also be noted that the presence of the distributions $\delta^{(1)}(t)$ and  $H(t)$ inside the left hand side parenthese 
in \eqref{eq:poft2} actually means the {\em removal} of these distributions from the left hand side expression.

The expression \eqref{eq:alltimebounds} generally provides an early-time (all-time) bound given that $p(t)*f(t)$ is unknown but $f(t)$, $\delta_tf(t)$ and $\delta_t^{-1}f(t)$ are known
as well as the coefficients $a_{-1}$, $b_{-1}$ and $b_1$.
It is noted that the case with $b_{-1}=a_{-1}$ is the trivial case when $p(s)=b_1s+a_{-1}s^{-1}$ and $p(t)*f(t)=b_1\delta_tf(t)+a_{-1}\delta_t^{-1}f(t)$.
In case we do not know $f(t)$, $\delta_tf(t)$ and $\delta_t^{-1}f(t)$ in explicit mathematical form but can estimate the
integral $B=\int_0^\infty f(\tau)\mrm{d}\tau$, we can exploit the inequality $\delta_t^{-1}f(t)\leq B$ 
extended directly on the right hand side of \eqref{eq:alltimebounds} to obtain a constant all-time bound valid for all $t>0$, see also \cite{Stumpf-Nordebo:2024a}.

We will finally present a simple early-time bound for a generalized step response which will become particularly useful together with a 
dielectric response (permittivity function) in what follows.
Here, we will use the following Laplace transforms
\begin{equation}\label{eq:gensteps}
\left\{\begin{array}{l}
f(s)=\frac{1/\tau}{s^2(s+1/\tau)}=\frac{\tau}{s+1/\tau}+\frac{1}{s^2}-\frac{\tau}{s}  \vspace{0.2cm} \\
sf(s)=\frac{1/\tau}{s(s+1/\tau)}=\frac{1}{s}-\frac{1}{s+1/\tau}  \vspace{0.2cm} \\
\frac{1}{s}f(s)=\frac{1/\tau}{s^3(s+1/\tau)}=\frac{\tau}{s(s+1/\tau)}+\frac{1}{s^3}-\frac{\tau}{s^2},
\end{array}\right.
\end{equation}
and employ \eqref{eq:alltimebounds} together with the corresponding time-domain expressions
\begin{equation}\label{eq:genstept}
\left\{\begin{array}{l}
f(t)=(\tau\eu^{-t/\tau}+t-\tau)H(t) \vspace{0.2cm} \\
\delta_t f(t)=(1-\eu^{-t/\tau})H(t) \vspace{0.2cm} \\
 \delta_t^{-1}f(t)=(\tau^2(1-\eu^{-t/\tau})+\frac{1}{2}t^2-\tau t)H(t).
\end{array}\right.
\end{equation}
Thus, the generalized step function is given by $\delta_t f(t)=(1-\eu^{-t/\tau})H(t)$ and which is now associated with the raise time $\tau>0$.
From the partial fractions in \eqref{eq:gensteps} it is readily seen that the case with $\tau=0$ is perfectly consistent with the case where
$\delta_t f(t)=H(t)$ is the ordinary unit step function, $f(t)=tH(t)$ is the ramp and $ \delta_t^{-1}f(t)=\frac{1}{2}t^2H(t)$.

\subsection{Late-time bounds}
The next useful possibility of exploiting sum rules for PR functions comes from the ramp response
\begin{equation}\label{eq:Cauerramp}
\frac{1}{s^2}p(s) =b_1s^{-1}+a_{-1}s^{-3}+\int_{\R\setminus\{0\}}\frac{1}{s(s^2+\xi^2)}\mrm{d}\beta(\xi),
\end{equation}
yielding the following inverse Laplace transform 
\begin{multline}\label{eq:rampoft}
\delta_t^{-2} p(t)=b_1H(t)+a_{-1}\frac{1}{2}t^2H(t) \\
+H(t)\int_{\R\setminus\{0\}}\frac{1}{\xi^2}\left(1-\cos(\xi t)\right)\mrm{d}\beta(\xi).
\end{multline}
Let us now furthermore assume that there exists an odd ordered low-frequency asymptotic expansion of order 1
\begin{equation}\label{eq:lowfrequency}
p(s)=\left\{\begin{array}{l}
a_{-1}s^{-1}+a_{1}s+o(s), \quad \textrm{as } s\toh 0 \vspace{0.2cm} \\
b_1s+o(s), \quad \textrm{as } s\toh\infty,
\end{array}\right.
\end{equation}
we have then the following sum rule \eqref{eq:Herglotz_SR1PR} for $n=2$
\begin{equation}
\int_{\R\setminus\{0\}}\frac{\mrm{d}\beta(\xi)}{\xi^2}=a_{1}-b_{1}.
\end{equation}
From the positivity of the measure it is concluded that $a_{1}-b_{1}\geq 0$ and where $a_{1}=b_{1}$ corresponds to the trivial case for which the positive measure $\beta$ is 
different from zero on $\R\setminus\{0\}$ only at a set of measure zero. 
From \eqref{eq:rampoft} follows then that
\begin{multline}
\pm\left(\delta_t^{-2} p(t)-b_1H(t)-a_{-1}\frac{1}{2}t^2H(t)\right) \\
=\pm H(t)\int_{\R\setminus\{0\}}\frac{1}{\xi^2}\left(1-\cos(\xi t)\right)\mrm{d}\beta(\xi) \\
\leq H(t)\int_{\R\setminus\{0\}}\frac{1}{\xi^2}2\mrm{d}\beta(\xi)=2(a_{1}-b_{1} )H(t),
\end{multline}
and hence
\begin{equation}\label{eq:rampoft2}
\left|\delta_t^{-2} p(t)-b_1H(t)-a_{-1}\frac{1}{2}t^2H(t)\right| 
\leq 2(a_{1}-b_{1} )H(t).
\end{equation}
It is noted that the trivial case with $a_{1}=b_{1}$ implies that $p(s)=b_1s+a_{-1}s^{-1}$ and 
$\delta_t^{-2} p(t)=b_1H(t)+a_{-1}\frac{1}{2}t^2H(t)$.

Finally, by assuming that there exists both a low-frequency as well as a high-frequency odd asymptotic expansion of order 1
\begin{equation}\label{eq:generalasymptoticsM1}
p(s)=\left\{\begin{array}{l}
a_{-1}s^{-1}+a_{1}s+o(s), \quad \textrm{as } s\toh 0 \vspace{0.2cm} \\
b_1s+b_{-1}s^{-1}+o(s^{-1}), \quad \textrm{as } s\toh\infty,
\end{array}\right.
\end{equation}
we can then combine the result \eqref{eq:alltimebounds} using the ramp $f(t)=tH(t)$ for which $f(s)=s^{-2}$ ($\tau=0$ in \eqref{eq:gensteps} and\eqref{eq:genstept})
together with \eqref{eq:rampoft2} to get the more general all-time bound
\begin{multline}\label{eq:combinedbound}
\left|\delta_t^{-2} p(t)-b_1H(t)-a_{-1}\frac{1}{2}t^2H(t)\right| \\
\leq
\left\{\begin{array}{ll}
(b_{-1}-a_{-1})\frac{1}{2}t^2H(t) & t\leq t_\mrm{c} \vspace{0.2cm} \\
2(a_{1}-b_{1} )H(t) &  t \geq t_\mrm{c}
\end{array}\right.
\end{multline}
and where the {\em corner time} $t_\mrm{c}$ is given by
\begin{equation}\label{eq:cornertime}
t_\mrm{c}=\sqrt{\frac{4(a_{1}-b_{1})}{b_{-1}-a_{-1}}},
\end{equation}
\cf also \cite{Stumpf-Nordebo:2024a}.

\section{Step response of a dielectric material}\label{sect:StepDielectric}
It is well known that the normalized dielectric constant (permittivity function) $\epsilon(s)$ of a passive material is associated with the positive real function
\begin{equation}
p(s)=s\epsilon(s)
\end{equation}
so that $p(t)=\delta_t\epsilon(t)$, \cf \cite{Gustafsson+Sjoberg2010a,Nedic+etal2019}. Thus, provided that the asymptotics \eqref{eq:generalasymptoticsM1} is valid
and the bounds in \eqref{eq:combinedbound} are applicable, we obtain an interesting time-domain bound for the unit step response of a dielectric material
involving a quadratic early-time bound and a constant late-time bound as
\begin{multline}\label{eq:combinedboundepsilon}
\left|\epsilon(t)*H(t)-b_1H(t)-a_{-1}\frac{1}{2}t^2H(t)\right| \\
\leq
\left\{\begin{array}{ll}
(b_{-1}-a_{-1})\frac{1}{2}t^2H(t) & t\leq t_\mrm{c} \vspace{0.2cm} \\
2(a_{1}-b_{1} )H(t) &  t \geq t_\mrm{c},
\end{array}\right.
\end{multline}
where we have used that $\delta_t^{-2}p(t)=\delta_t^{-1} \epsilon(t)=\epsilon(t)*H(t)$
and where the corner time is given by \eqref{eq:cornertime}.

Given that it is only the high-frequency asymptotics \eqref{eq:highfrequency} that can be confirmed we can only refer to the following early-time bound 
\begin{equation}\label{eq:etstepepsilon}
\left|\epsilon(t)*H(t)-b_1H(t)-a_{-1}\frac{1}{2}t^2H(t)\right|\leq (b_{-1}-a_{-1})\frac{1}{2}t^2H(t),
\end{equation}
but which is valid for all $t$ and asymptotically accurate as $t\rightarrow 0+$. However, in this case we may also incorporate the more general bound
\eqref{eq:alltimebounds} with $f(t)$ defined by \eqref{eq:gensteps} and \eqref{eq:genstept} yielding 
\begin{multline}\label{eq:genstepepsilon}
\left| \epsilon(t)*\delta_tf(t)-b_1\delta_tf(t)-a_{-1}\delta_t^{-1}f(t)\right| \\
\leq (b_{-1}-a_{-1})\delta_t^{-1}f(t)
\end{multline}
where $\delta_t f(t)=(1-\eu^{-t/\tau})H(t)$ is the generalized step function with raise time $\tau$ and $\delta_t^{-1}f(t)=(\tau^2(1-\eu^{-t/\tau})+\frac{1}{2}t^2-\tau t)H(t)$.

Let us now investigate the asymptotic properties of some standard dielectric models 
to see whether they qualify for the bounds given by \eqref{eq:combinedboundepsilon}, \eqref{eq:etstepepsilon} and \eqref{eq:genstepepsilon}.
The corresponding physical bounds are then valid for all passive materials sharing the same basic first order asymptotics as the standard model under consideration.

\subsection{Standard conductivity model}
The standard conductivity model is commonly used to model the electrical conductivity of a solid or a liquid and is given by
\begin{equation}
\epsilon(s)=\epsilon_\infty+\frac{\sigma}{s\epsilon_0},
\end{equation}
where $\epsilon_\infty>0$ is the optical response, $\sigma\geq 0$ the conductivity and $\epsilon_0$ the permittivity of free space (vacuum). 
The parameter $\epsilon_\infty$ is usually taken to be $\epsilon_\infty=1$.
The corresponding PR function and its first order asymptotics are given by
\begin{equation}\label{eq:pconductivity}
p(s)=\epsilon_\infty s+\frac{\sigma}{\epsilon_0}=\left\{\begin{array}{l}
\frac{\sigma}{\epsilon_0}+\epsilon_\infty s+o(s) \quad \textrm{as } s\toh 0 \vspace{0.2cm} \\
\epsilon_\infty s+\frac{\sigma}{\epsilon_0}+o(s^{-1}) \quad \textrm{as } s\toh\infty.
\end{array}\right.
\end{equation}
We can see that the requirements given by \eqref{eq:generalasymptoticsM1} are not satisfied here
and the bound given by \eqref{eq:combinedboundepsilon} can not be applied.
In fact, neither of the high- or low-frequency asymptotics of odd order 1 defined in \eqref{eq:highfrequency} or \eqref{eq:lowfrequency} are satisfied here
and hence neither of the corresponding early- and late-time bounds given by \eqref{eq:alltimebounds} or \eqref{eq:rampoft2} can be applied.

\subsection{Debye model}
The Debye model is commonly used to model the response of dielectric media with permanent molecular dipole moment (\eg water or other polar liquids) and is given by
\begin{equation}\label{eq:Debyemodel}
\epsilon(s)=\epsilon_{\infty}+\frac{\epsilon_{\rm s}-\epsilon_{\infty}}{1+s\tau_\mrm{r}},
\end{equation}
where $\epsilon_\infty>0$ is the optical response, $\epsilon_{\mrm s}>0$ the static response and $\tau_\mrm{r}>0$ the relaxation time.
The corresponding PR function and its first order asymptotics are given by
\begin{multline}\label{eq:pDebye}
p(s)=s\epsilon_{\infty}+s\frac{\epsilon_{\rm s}-\epsilon_{\infty}}{1+s\tau_\mrm{r}} \\
=\left\{\begin{array}{l}
\epsilon_{\rm s}s+o(s) \quad \textrm{as } s\toh 0 \vspace{0.2cm} \\
\epsilon_\infty s+\frac{\epsilon_{\rm s}-\epsilon_{\infty}}{\tau_\mrm{r}}-\frac{\epsilon_{\rm s}-\epsilon_{\infty}}{\tau_\mrm{r}^2}s^{-1}+o(s^{-1}) \quad \textrm{as } s\toh\infty.
\end{array}\right.
\end{multline}
We can see that the requirements given by \eqref{eq:generalasymptoticsM1} are not satisfied here
and the bound given by \eqref{eq:combinedboundepsilon} can not be applied.
However, the low-frequency asymptotics of \eqref{eq:lowfrequency} is valid and 
we do have the late-time bound given by \eqref{eq:rampoft2} where $a_{-1}=0$,  $a_1=\epsilon_{\rm s}$ and $b_1=\epsilon_\infty$
and where the static permittivity is always greater than the optical response, \ie $\epsilon_{\rm s}\geq \epsilon_\infty$.
The corresponding TD bound is given by
\begin{equation}\label{eq:Debye}
\left| \epsilon(t)*H(t)-\epsilon_\infty H(t)\right|\leq 2(\epsilon_{\rm s}-\epsilon_\infty )H(t),
\end{equation}
and where we have again used that $\delta_t^{-2}p(t)=\delta_t^{-1} \epsilon(t)=\epsilon(t)*H(t)$. Notably, if $\epsilon_{\rm s}=\epsilon_\infty$ then
$\epsilon(t)=\epsilon_\infty\delta(t)$ and $\epsilon(s)=\epsilon_\infty$.

\subsection{Lorentz model}
The Lorentz model is commonly used to model the dielectric response of solids and gases with bound charges, and is given by
\begin{equation}
\epsilon(s)=\epsilon_{\infty}+\frac{\omega_{\mrm p}^2}{s^2+s\nu+\omega_0^2}
\end{equation}
where $\epsilon_\infty>0$ is the optical response, $\omega_{\rm p}>0$ the plasma frequency, $\omega_0>0$ the resonance frequency and 
$\nu\geq 0$ the collision frequency. 
The case with $\omega_0=0$ gives the Drude model which is treated below.
The corresponding time-domain impulse response and unit step response for an underdamped system where $\omega_0> \nu/2$ are given by
\begin{equation}\label{eq:TDLorentz}
\left\{\begin{array}{lcl}
\epsilon(t) & = & \epsilon_{\infty}\delta(t)+\frac{\omega_{\rm p}^2}{\nu_0}\eu^{-\nu t/2}\sin(\nu_0t)H(t), \vspace{0.2cm} \\
\epsilon(t)*H(t) & = & \epsilon_{\infty}H(t)+\frac{\omega_{\rm p}^2}{\omega_0^2}\left(1-\eu^{-\nu t/2}\left(\cos(\nu_0t) \right.\right. \vspace{0.2cm} \\
    &      &   \left. \left.  +\frac{\nu}{2\nu_0}\sin(\nu_0t)\right)\right)H(t),
\end{array}\right.
\end{equation}
where $\nu_0=\sqrt{\omega_0^2-\nu^2/4}>0$.
The corresponding PR function and its first order asymptotics are given by
\begin{multline}\label{eq:pLorentz}
p(s)=s\epsilon_{\infty}+\frac{s\omega_{\rm p}^2}{s^2+s\nu+\omega_0^2} \\
=\left\{\begin{array}{l}
\epsilon_{\rm s}s+o(s) \quad \textrm{as } s\toh 0, \vspace{0.2cm} \\
\epsilon_\infty s+\omega_{\rm p}^2s^{-1}+o(s^{-1}) \quad \textrm{as } s\toh\infty,
\end{array}\right.
\end{multline}
where $\epsilon_{\mrm s}=\epsilon_{\infty}+\omega_{\rm p}^2/\omega_{0}^2$ is the static permittivity. 
We can see that the requirements given by \eqref{eq:generalasymptoticsM1} are satisfied here with
\begin{equation}
\left\{\begin{array}{ll}
a_{-1}=0 & a_1=\epsilon_\mrm{s} \vspace{0.2cm} \\
b_{1}=\epsilon_\infty & b_{-1}=\omega_\mrm{p}^2.
\end{array}\right.
\end{equation}
It is also noticed that the first order asymptotic parameters above are independent of the loss parameter $\nu$ and where $0\leq \nu < 2\omega_0$.

\begin{figure}
\begin{center}
\includegraphics[width=0.48\textwidth]{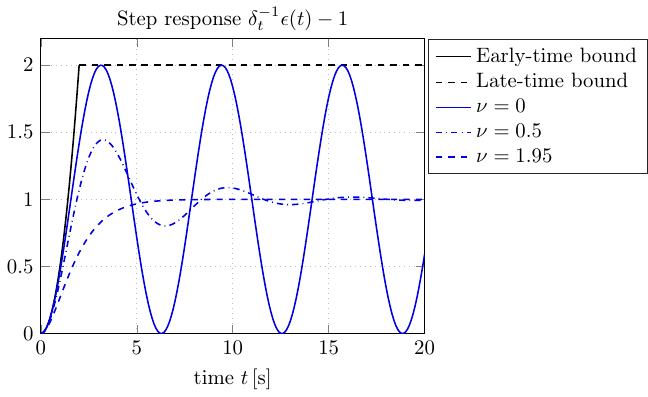}
\end{center}
\vspace{-5mm}
\caption{Early- and late-time bounds for the unit step response of a dielectric constant with first order asymptotics given by \eqref{eq:pLorentz} 
and a comparison with the actual response of the Lorentz model \eqref{eq:TDLorentz} with $\epsilon_\infty=1$, $\omega_0=\omega_\mrm{p}=1$
and $\nu\in\{0, 0.5, 1.95\}$. The bounds are in black color and the Lorentz responses are in blue color.
}
\label{fig:Lorentzfig}
\end{figure}

We may now consider a general passive dielectric material with the same first order asymptotics as in \eqref{eq:pLorentz} and conclude that the bound \eqref{eq:combinedboundepsilon} 
is valid. The bound is given here explicitly as
\begin{equation}\label{eq:combinedboundepsilonLor}
\left| \epsilon(t)*H(t)-\epsilon_\infty H(t)\right|\leq
\left\{\begin{array}{ll}
\omega_\mrm{p}^2\frac{1}{2}t^2H(t) & t\leq t_\mrm{c} \vspace{0.2cm} \\
2(\epsilon_\mrm{s}-\epsilon_\infty)H(t) 
&  t \geq t_\mrm{c},
\end{array}\right.
\end{equation}
and where the corner time \eqref{eq:cornertime} is given by
\begin{equation}\label{eq:cornertimeLor}
t_\mrm{c}
=\sqrt{\frac{4(\epsilon_\mrm{s}-\epsilon_\infty)}{\omega_\mrm{p}^2}}
=\sqrt{\frac{4\omega_{\rm p}^2/\omega_{0}^2}{\omega_\mrm{p}^2}}=\frac{2}{\omega_0}.
\end{equation}
A numerical example is given in Fig.~\ref{fig:Lorentzfig}. The Lorentz model is implemented here with $\epsilon_\infty=1$, $\omega_0=\omega_\mrm{p}=1$
and $0\leq \nu < 2$. 

It is noted that the bound is tight in the loss-less case when $\nu=0$.
When losses are non-zero and $\nu>0$, we can see that $\left| \epsilon(t)*H(t)-\epsilon_\infty H(t)\right|\rightarrow \epsilon_\mrm{s}-\epsilon_\infty=\omega_{\rm p}^2/\omega_{0}^2$
as $t\rightarrow\infty$ in accordance with the final value theorem $\lim_{t\rightarrow\infty} \delta_t^{-1} \epsilon(t)=\lim_{s\toh 0}\epsilon(s)=\epsilon_\mrm{s}$ where $\epsilon_\mrm{s}$ is the static permittivity.
Obviously, a positive combination of multiple resonances can be treated similarly. It is also interesting to observe that the squared plasma frequency $\omega_{\rm p}^2$
could potentially be determined from accurate measurements of the early-time asymptotic quadratic response as of \eqref{eq:combinedboundepsilonLor} for $t\ll t_\mrm{c}$.

\subsection{Drude model}
The Drude model is used to model the conduction of charges in metals and is a special case of the
Lorent'z model with resonance frequency $\omega_0=0$. Here
\begin{equation}
\epsilon(s)=\epsilon_{\infty}+\frac{\omega_\mrm{p}^2}{s(s+\nu)}
\end{equation}
where $\epsilon_\infty>0$ is the optical response, 
$\omega_{\rm p}>0$ the plasma frequency and $\nu>0$ the collision frequency.
In this case it is straightforward to derive the following generalized step response by using standard Laplace transform methods yielding
\begin{multline}\label{eq:Druderesponse1}
\epsilon(t)*(1-\eu^{-t/\tau})H(t)=\epsilon_\infty(1-\eu^{-t/\tau})H(t) \\
+\frac{\omega_\mrm{p}^2}{\nu^2}\left( \frac{\tau^2\nu^2\eu^{-t/\tau}-\eu^{-\nu t}}{\tau\nu-1} +\nu t-\tau\nu -1\right)H(t)
\end{multline}
where $\tau>0$ is the raise time of the generalized step function. In case the excitation is an ideal unit step function, we can set $\tau=0$ and obtain
\begin{equation}\label{eq:Druderesponse2}
\epsilon(t)*H(t)=\epsilon_\infty H(t) 
+\frac{\omega_\mrm{p}^2}{\nu^2}\left( \eu^{-\nu t} +\nu t -1\right)H(t).
\end{equation}

The corresponding PR function and its first order asymptotics are given by
\begin{multline}\label{eq:pDrude}
p(s)=s\epsilon_{\infty}+\frac{\omega_\mrm{p}^2}{s+\nu} \\
=\left\{\begin{array}{l}
\frac{\omega_\mrm{p}^2}{\nu}+(\epsilon_\infty-\frac{\omega_\mrm{p}^2}{\nu^2})s+o(s) \quad \textrm{as } s\toh 0 \vspace{0.2cm} \\
\epsilon_\infty s+\omega_{\rm p}^2s^{-1}+o(s^{-1}) \quad \textrm{as } s\toh\infty.
\end{array}\right.
\end{multline}
We can see that the requirements given by \eqref{eq:generalasymptoticsM1} are not satisfied here
and the bound given by \eqref{eq:combinedboundepsilon} can not be applied.
However,  the high-frequency asymptotics \eqref{eq:highfrequency} is valid and
we do have the early-time bounds given by \eqref{eq:alltimebounds} where $b_1=\epsilon_\infty$, $b_{-1}=\omega_\mrm{p}^2$ and $a_{-1}=0$.
The corresponding early-time bound \eqref{eq:genstepepsilon} is then given by
\begin{equation}\label{eq:earlytimeDrude1}
\left| \epsilon(t)*\delta_tf(t)-\epsilon_\infty\delta_tf(t)\right| \leq \omega_\mrm{p}^2 \delta_t^{-1}f(t).
\end{equation}
Thus, in the case when the raise time $\tau>0$, we can write the bound explicitly as
\begin{multline}\label{eq:earlytimeDrude2}
\left| \epsilon(t)*(1-\eu^{-t/\tau})H(t)-\epsilon_\infty(1-\eu^{-t/\tau})H(t)\right| \\
\leq  \omega_\mrm{p}^2\left(\tau^2(1-\eu^{-t/\tau})+\frac{1}{2}t^2-\tau t\right)H(t),
\end{multline}
and where $\delta_tf(t)$ and $\delta_t^{-1}f(t)$ have been inserted according to \eqref{eq:genstept}.
In case the input is the standard unit step function with raise time $\tau=0$, we have instead
$\delta_t f(t)=H(t)$ for which $\delta_t^{-1}f(t)=\frac{1}{2}t^2H(t)$. The bound in \eqref{eq:earlytimeDrude2} can now
be compared with the corresponding Drude responses in \eqref{eq:Druderesponse1} and \eqref{eq:Druderesponse2}.

\begin{figure}
\begin{center}
\includegraphics[width=0.35\textwidth]{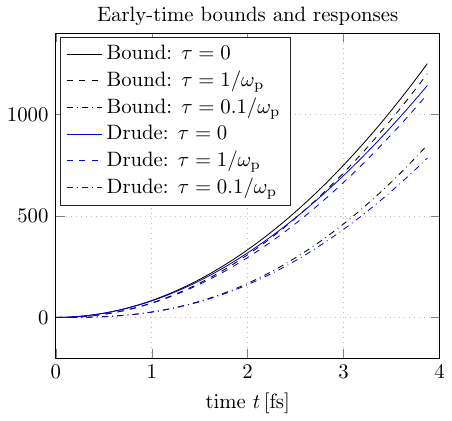}
\end{center}
\vspace{-5mm}
\caption{Early-time bounds for the generalized step response \eqref{eq:earlytimeDrude2} with pulse raise time $\tau$
and asymptotics given by \eqref{eq:pDrude} and a comparison with the actual response
of the Drude model \eqref{eq:Druderesponse1} for gold (Au) with plasma frequency $\omega_\mrm{p}$ and losses $\nu$
according to the free electron model of Olmon et al\cite{Olmon+etal2012}.
The plot is made in femtoseconds for $t\in[0,50/\omega_\mrm{p}]$. The bounds are in black color and the Drude responses are in blue color.
}
\label{fig:BBfig3}
\end{figure}

\begin{figure}
\begin{center}
\includegraphics[width=0.35\textwidth]{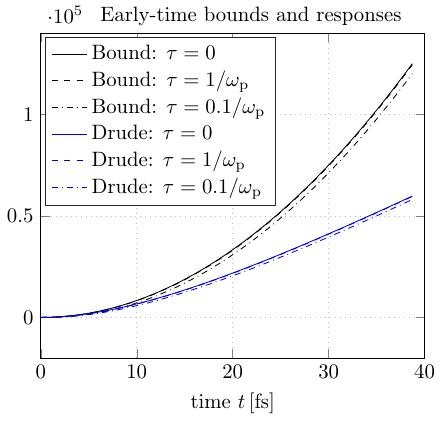}
\end{center}
\vspace{-5mm}
\caption{Same plot as in Fig.~\ref{fig:BBfig3}, only that the
plot is made here for the larger time range $t\in[0,500/\omega_\mrm{p}]$. The bounds are in black color and the Drude responses are in blue color.
}
\label{fig:BBfig4}
\end{figure}

In Figs.~\ref{fig:BBfig3} and \ref{fig:BBfig4} are shown
the early-time bounds for the generalized step response \eqref{eq:earlytimeDrude2} 
with asymptotics given by \eqref{eq:pDrude} and a comparison with the actual response
of the Drude model \eqref{eq:Druderesponse1} for gold (Au) according to the free electron model of Olmon et al\cite{Olmon+etal2012}.
Here, $\epsilon_\infty=1$,  $\omega_\mrm{p}=1.29\cdot 10^{16}\unit{s^{-1}}$ ($\hbar\omega_\mrm{p}=8.5\unit{eV}$)
and $\nu=7.14\cdot 10^{13}\unit{s^{-1}}$ ($\hbar\nu = 0.047\unit{eV}$).
The pulse raise time is $\tau\in\{0,1/\omega_\mrm{p},0.1/\omega_\mrm{p}\}$.

As can be seen in Figs.~\ref{fig:BBfig3} and \ref{fig:BBfig4} the actual Drude responses are rather close to the corresponding  upper bounds
in the shorter time range up to 4\unit{fs} ($\sim 50/\omega_\mrm{p}$) but starts to deviate significantly in the larger time range up to 40\unit{fs} ($\sim 500/\omega_\mrm{p}$) 
due to the quadratic nature of the early-time bound. The significance of the bound in this case lies in the fact that the material behaves as
a conductor at low frequencies (large time scales) and as a loss-less dielectric at high frequencies (short time scales). To this end, the early-time bound can
be used to quantify at which time scales and pulse raise times the material response will behave adequately (or inadequately) as a conductor or as an insulator.
This will then be valid for any passive material sharing the same first order asymptotics as the particular Drude model under consideration.
The next model to investigate under these same circumstances is the Brendel-Bormann model.

\subsection{The Brendel Bormann model}
A widely accepted non-rational model for the dielectric response of metals and amorphous solids
is given by the Brendel-Bormann (BB) model \cite{Brendel+Bormann1992,Rakic+etal1998}. 
Here, the electric susceptibility function of a single resonance ($j=1,\ldots,k$) is given by a Gaussian distribution of Lorentzian oscillators as
\begin{equation}\label{eq:BBchimodel}
\chi_j(\omega)=\frac{1}{\sqrt{2\pi}\sigma_j}\int_{-\infty}^{\infty}\eu^{-(x-\omega_j)^2/2\sigma_j^2}\frac{\omega_{\mrm{p}j}^2}{x^2-\omega^2-\iu\omega\nu_j}\mrm{d}x,
\end{equation}
where $\omega_j$ is the resonance frequency, $\omega_{\mrm{p}j}$ the plasma frequency of the Lorentzian, $\nu_j$ the line width of the Lorentzian
and $\sigma_j^2$ the variance of the Gaussian distribution. The total dielectric function is then modeled as
\begin{equation}\label{eq:BBepsmodel}
\epsilon(\omega)=1-\frac{\omega_{\mrm{p}0}^2}{\omega(\omega+\iu\nu_0)}+\sum_{j=1}^k \chi_j(\omega),
\end{equation}
where the second term is an ordinary Drude model with parameters $\omega_{\mrm{p}0}$ and $\nu_0$, \cf \cite[Eq. (11)]{Rakic+etal1998}.
It is noted that there is no static permittivity associated with neither of the ordinary Drude model nor with the Brendel Bormann model
as $\epsilon(\omega)$ is singular at $\omega=0$.

The models in \eqref{eq:BBchimodel} and \eqref{eq:BBepsmodel} are given here in terms of the Fourier-Laplace transform where
the Laplace variable is $s=-\iu\omega$. Thus, the adequate Herglotz function here is $h(\omega)=\omega\epsilon(\omega)$
and the corresponding PR function is $p(s)=-\iu h(\omega)=s\epsilon(s)$, as above. It is clear that \eqref{eq:BBepsmodel} generates a Herglotz function
$\omega\epsilon(\omega)$ since $\omega\chi_j(\omega)$ is in the positive cone generated by the Lorentzian Herglotz functions $\omega/(x^2-\omega^2-\iu\omega\nu_j)$
for all $x\in\R$ as in \eqref{eq:BBchimodel}.

Let us now investigate the asymptotic properties of the Herglotz function $h(\omega)=\omega\epsilon(\omega)$ given by \eqref{eq:BBepsmodel}.
For this purpose we may now exploit the fact that $\chi_j(\omega)$ can be expressed as
\begin{equation}\label{eq:BBchimodelFadd}
\chi_j(\omega)=\frac{\omega_{\mrm{p}j}^2\sqrt{\pi}}{2\sqrt{2}\alpha_j\sigma_j}\iu\left(w\left(\frac{\alpha_j-\omega_j}{\sqrt{2}\sigma_j}\right) 
+ w\left(\frac{\alpha_j+\omega_j}{\sqrt{2}\sigma_j}\right) \right),
\end{equation}
where $w(\cdot)$ is the Faddeeva function and $\alpha_j=\sqrt{\omega^2+\iu\omega\nu_j}$ where $\Im\{\alpha_j\}\geq 0$, \cf \cite{Brendel+Bormann1992,Rakic+etal1998}.
In fact, a representation of the BB model based on \eqref{eq:BBchimodelFadd} is tractable for numerical reasons as well as for analytical purposes.
There is a vast literature on the development of fast and accurate numerical methods for the computation of the Faddeeva function, 
see \eg \cite{Armstrong1967,Humlicek1982,Imai+etal2010,Kuntz1997,Schreier2011,Letchworth+Benner2007,Abrarov+Quine2011,Nordebo2021a}, only to mention a few,
and a typical application is within quantitative spectroscopy, see \eg \cite{Tennyson+etal2014} with references.
The analytical properties of the Faddeeva function are furthermore well established and readily applicable as will be demonstrated below.
In particular, the Faddeeva function is defined by $w(z)=\eu^{-z^2}\mrm{erfc}(-\iu z)$ where $\mrm{erfc}(z)$ is the complementary error function
$\mrm{erfc}(z)=\frac{2}{\sqrt{\pi}}\int_z^\infty\eu^{-t^2}\mrm{d}t$ \cf \cite[Eq.~(7.2.1)-(7.2.3)]{Olver+etal2010}. 
The Faddeeva function also has an integral representation given by
\begin{equation}\label{eq:Faddeevaint}
w(z)=\frac{\iu}{\pi}\int_{-\infty}^\infty\frac{\eu^{-\xi^2}}{z-\xi}\mrm{d}\xi,
\end{equation}
which is valid for $\Im\{z\}>0$ and which is showing that $\iu w(z)$ is a Herglotz function and $\Re\{w(z)\}>0$ for $\Im\{z\}>0$, \cf \cite[Eq.~7.1.4]{Abramowitz+Stegun1970}. 
The small- and large argument asymptotics of $w(z)$ are furthermore given by
\begin{equation}\label{eq:Faddeevaasy}
w(z)=\left\{\begin{array}{l}
1+\iu\frac{2}{\sqrt{\pi}}z+{\cal O}\{z^2\}, \quad \textrm{as } z \rightarrow 0, \vspace{0.2cm} \\
 \iu\frac{1}{\sqrt{\pi}}\frac{1}{z}+{\cal O}\{\frac{1}{z^3}\}, \quad \textrm{as } z \rightarrow \infty,
\end{array}\right.
\end{equation}
where the first expression is a Taylor series expansion at $z=0$ and the second expansion is valid for $-\pi/4<\arg(z)<5\pi/4$,
 \cf \cite[Eq.~(7.6.3)]{Olver+etal2010} and \cite[Eq.~(7.12.1)]{Olver+etal2010}, respectively. By carefully investigating
 the model \eqref{eq:BBchimodelFadd} in view of the asymptotics given by \eqref{eq:Faddeevaasy} as well as the factor 
 \begin{equation}
\frac{ \omega}{\alpha_j}=\frac{\omega}{\sqrt{\omega^2+\iu\omega\nu_j}},
 \end{equation}
 it is readily found that
\begin{equation}
\omega \chi_j(\omega)=
 \left\{\begin{array}{l}
o(\omega^{-1}), \quad \textrm{as } \omega \rightarrow 0, \vspace{0.2cm} \\
-\omega_{\mrm{p}j}^2\omega^{-1}+o(\omega^{-1}), \quad \textrm{as } \omega \rightarrow \infty.
\end{array}\right.
 \end{equation}
In fact, at low frequencies we can see that $\omega \chi_j(\omega)=C\sqrt{\omega}+o(\sqrt{\omega})$ where $C$ is a constant, indicating that the only
useful information that we can retrieve from the low-frequency asymptotics here is that the coefficient $a_{-1}=0$. 
For comparison, it is seen that the Drude term is $\omega \chi_0(\omega)=\iu\omega_{\mrm{p}0}^2/\nu_0+o(1)=o(\omega^{-1})$ for small $\omega$, also
of {\em odd} asymptotic order $-1$.
Hence, similar to the Drude model our focus must be solely on the high-frequency asymptotics \eqref{eq:highfrequency} 
together with the early-time bounds given by \eqref{eq:etstepepsilon} and \eqref{eq:genstepepsilon}.

From the analysis above follows that the asymptotics of the positive real function $p(s)=s\epsilon(s)$ corresponding to the BB model given by \eqref{eq:BBepsmodel} is given by
\begin{equation}\label{eq:pBB}
p(s)= \left\{\begin{array}{l}
o(s^{-1}), \quad \textrm{as } s \rightarrow 0, \vspace{0.2cm} \\
s+\displaystyle\sum_{j=0}^k\omega_{\mrm{p}j}^2s^{-1}+o(s^{-1}), \quad \textrm{as } s \rightarrow \infty,
\end{array}\right.
\end{equation}
and where we have used again that $p(s)=-\iu h(\omega)$ and $s=-\iu\omega$.
We can now conclude that the early-time bounds \eqref{eq:etstepepsilon} and \eqref{eq:genstepepsilon} are applicable with
$a_{-1}=0$, $b_1=1$ and $b_{-1}=\omega_\mrm{p}^2$ is the equivalent plasma frequency where
\begin{equation}\label{eq:omegapBB}
\omega_\mrm{p}^2=\sum_{j=0}^k\omega_{\mrm{p}j}^2.
\end{equation}
The corresponding early-time bound \eqref{eq:genstepepsilon} is then given explicitly as in \eqref{eq:earlytimeDrude2} 
where $\epsilon_\infty=1$, $\omega_\mrm{p}^2$ is given by \eqref{eq:omegapBB} and where \eqref{eq:genstept} has been incorporated again.

The optical constants of 11 metals have been modeled with 
Brendel-Bormann parameters $(\omega_{\mrm{p}j},\sigma_j,\omega_j,\nu_j)$ and fitted to experimental data in \cite[Eq.~(11) with parameters from Table 1 and Table 3]{Rakic+etal1998}.
The corresponding plasma frequencies $\omega_\mrm{p}$ and characteristic times $1/\omega_\mrm{p}$ are calculated here according to \eqref{eq:omegapBB}
and summarized in Tab.~\ref{tab:metals} below. As we can see here, the variation in plasma frequency among the various metals is not very large.

\begin{figure}
\begin{center}
\includegraphics[width=0.35\textwidth]{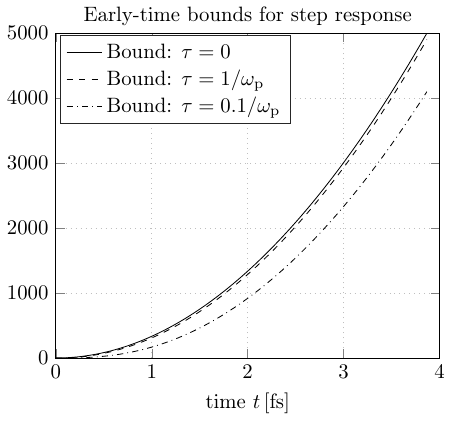}
\end{center}
\vspace{-5mm}
\caption{
Early-time bounds for the generalized step response \eqref{eq:earlytimeDrude2} with pulse raise time $\tau$
and asymptotics given by \eqref{eq:pBB}. 
Here, the equivalent plasma frequency $\omega_\mrm{p}$ is for gold (Au) according to the Brendel Bormann model \cite{Rakic+etal1998} summarized in Tab.~\ref{tab:metals}.
The plot is made in femtoseconds for $t\in[0,100/\omega_\mrm{p}]$.
}
\label{fig:BBfig2}
\end{figure}

In Fig.~\ref{fig:BBfig2} is illustrated the early-time bounds \eqref{eq:earlytimeDrude2} with asymptotics given by \eqref{eq:pBB}
according to the Brendel Bormann model of gold (Au) and where the equivalent plasma frequency is 
$\omega_\mrm{p}=2.58\cdot 10^{16}\unit{s^{-1}}$ ($\hbar\omega_\mrm{p}=17.0\unit{eV}$). 
Thus, the physical bound illustrated in Fig.~\ref{fig:BBfig2} is now valid for any passive dielectric media (such as gold) sharing
the same plasma frequency and first order asymptotics as the actual BB model under consideration. 
It would furthermore be expected that the BB model for gold would be reasonable close to the corresponding upper bounds in this very short-time interval under consideration,
similarly as with the Drude model for gold as illustrated in Fig.~\ref{fig:BBfig3}.

\setlength{\tabcolsep}{1pt}
\begin{table}[htb]\scriptsize\centerline{\renewcommand{\arraystretch}{1.5}
\begin{tabular}{||l||c|c|c|c|c|c|c|c|c|c|c|}
\cline{1-12} 
$\omega_\mrm{p}\backslash$ Xy &   Ag & Au & Cu & Al & Be & Cr & Ni & Pd & Pt & Ti & W     \\
\cline{1-12} $\hbar\omega_\mrm{p}\unit{[eV]} $   & 21.2  &  17.0  & 14.4 &  14.9 &  17.3 &  13.9  & 17.9  & 13.4 &  19.1 &   8.3 &  22.9         \\
\cline{1-12}  $1/\omega_\mrm{p}\unit{[as]}$  & 31.1 &  38.7  & 45.7 &  44.0 &  38.1  & 47.3  & 36.8  & 49.0 & 34.5  & 79.7  & 28.7          \\
\cline{1-12} 
\end{tabular}}
\vspace{3mm}
\caption{Equivalent plasma frequency $\hbar\omega_\mrm{p}$ for 11 metals in units of electronvolt\unit{(eV)} and the corresponding characteristic times $1/\omega_\mrm{p}$ in units of attoseconds\unit{(as)}
retrieved from the Brendel-Bormann models given by \cite{Rakic+etal1998}.}\label{tab:metals}
\end{table}

\section{Summary and conclusions}\label{sect:Summary}
Physical limitations on the time-domain response of a passive system has been presented in this paper.
The theory is based on Cauer's representation of an arbitrary Positive Real (PR) function together with associated sum rules
and exploits the unilateral Laplace transform to derive rigorous bounds directly in the time-domain.
The advantage of this approach is the ease by which rigorous physical bounds can be derived by exploiting the 
integral representation, its positive generating measure and associated sum rules. 
The method is however limited to PR functions having some odd ordered low- and/or high-frequency asymptotic expansion
for which the required sum rule exists. Hence, this field will be open to explore other subclasses of linear, time-invariant and casual systems beyond passive systems,
as suggested in\cite{Stumpf-Nordebo:2024a,Stumpf-Nordebo:2024b}.


\appendix

\subsection{Basic properties of Positive Real functions}\label{sect:PRbasics}
The set of Positive Real (PR) functions $\{p(s)\}$ is equivalent to the set of symmetric Herglotz functions $\{h(z)\}$ via the transformation
$p(s)=-\iu h (\iu s)$ where $s=-\iu z$. Their basic properties can therefore be deduced from one another based on an extensive literature found in
\eg \cite{Zemanian1965,Kac+Krein1974,Akhiezer1965,Nussenzveig1972,Gesztesy+Tsekanovskii2000,Nedic+etal2019} with references, and where we will employ
here in particular the survey given in \cite{Nedic+etal2019}. For convenience, it is practical here to set $z=x+\iu y$ and $s=\sigma-\iu\omega$ so that 
$\omega=x$ (frequency) and $\sigma=y$ (damping, or loss factor). The more conventional definition for the Laplace variable
$s=\sigma+\ju\omega$ is then obtained simply by making the substitution $\iu=-\ju$.

A PR function is a holomorphic function defined on the open right half-plane $\C_+=\{s\in\C|\Re\{s\}>0\}$ 
where its real part is non-negative, \ie $\Re\{p(s)\}\geq 0$ for $s\in\C_+$ and which satisfies the following symmetry
\begin{equation}\label{eq:Herglotz_symmetryPR}
p(s)=p^*(s^*),
\end{equation}
\cf \eg \cite[Definition 20.3]{Nedic+etal2019} and \cite[Chapt.~10.4]{Zemanian1965}. 
Any PR function is uniquely given by Cauer's representation \cite[Chapt. 10.5]{Zemanian1965}
\begin{equation}\label{eq:Cauer}
p(s) =bs+\int_{-\infty}^{\infty}\frac{s}{s^2+\xi^2}\mrm{d}\beta(\xi),
\end{equation}
where $b\geq 0$ and the positive Borel measure $\beta(\xi)$ 
is the same as for the corresponding Herglotz function \cite[Eq.~(20.13)]{Nedic+etal2019} with growth condition
\begin{equation}\label{eq:growthcond}
\int_{-\infty}^{\infty}\frac{\mrm{d}\beta(\xi)}{1+\xi^2}<\infty.
\end{equation}
The constant $b$ is determined by
\begin{equation}
b=\lim_{s\toh\infty}\frac{p(s)}{s}
\end{equation}
where the non-tangential limit is taken in the right half-plane 
($|s|\to \infty$ in the Stoltz cone $\varphi-\pi/2\leq\arg s \leq\pi/2-\varphi$ for any $\varphi\in(0,\pi/2]$).
The positive measure $\beta$ is furthermore uniquely determined by the PR function (Herglotz function $h(z)=\iu p(s)$)
from the Stieltjes inversion formula, see \cite{Kac+Krein1974,Gesztesy+Tsekanovskii2000}.
In particular, in the case when the measure is absolutely continuous we may write $\mrm{d}\beta(\xi)=\beta^\prime(\xi)\mrm{d} \xi$ where $\beta^\prime(\xi)$ is the density
of the measure and where 
\begin{equation}\label{eq:betaprime}
\beta^\prime(\xi)=\frac{1}{\pi}\lim_{\sigma\rightarrow 0+}\Re\{p(\sigma-\iu \xi)\}.
\end{equation}
It is noted that the measure is even and we have that $\mrm{d}\beta(-\xi)=-\mrm{d}\beta(\xi)$ and thus $\beta^\prime(-\xi)=\beta^\prime(\xi)$.
For point masses we have $\beta(\{\xi_0\})=\beta(\{-\xi_0\})$.

It is readily seen (by using residue calculus) that a real constant $p(s)=C$ with $C>0$ can be
generated by the constant measure $\mrm{d}\beta(t)=\frac{1}{\pi}C\mrm{d} t$.
It follows directly from the symmetry requirement \eqref{eq:Herglotz_symmetryPR} (as well as from the representation \eqref{eq:Cauer})
that $p(s)$ is real valued for real valued $s$. It can furthermore be shown that $\Re\{p(s)\}>0$ for $s\in\C_+$ unless $p(s)\equiv 0$.
Thus, it is perfectly safe (except for the trivial case $p(s)\equiv 0$) to generate new PR functions by inversion $1/p(s)$ as well as by composition $p_1(p_2(s))$ where both $p_1$ and $p_2$ are PR functions.

It can be shown that the measure $\beta$ has a point mass at the point $\xi_0\in\R$ if and only if the limit
\begin{equation}
\beta(\{\xi_0\})=\lim_{s\toh \iu\xi_0}(s-\iu\xi_0)p(s)>0.
\end{equation}
A simple example is $\mrm{d}\beta(\xi)=c\delta(\xi)\mrm{d} \xi$ generating the PR function $p(s)=\frac{c}{s}$
where $c=\beta(\{0\})>0$ and $\delta(\xi)$ is the Dirac delta function.

For an asymptotic expansion of the form $p(s)\sim \sum_n c_n s^n+o(\cdot)$ (either for $s\toh 0$ or $s\toh \infty$)
it is readily seen that the  symmetry  \eqref{eq:Herglotz_symmetryPR} implies that all coefficients $c_n$ must be real valued.
The relationship between the corresponding coefficients for a symmetric Herglotz function with $h(z)\sim \sum_n \tilde{c}_n z^n+o(\cdot)$, 
is thus given by $c_{n}=-\iu^{n+1}\tilde{c}_{n}$. Hence, with $n$ even we have $\tilde{c}_{n-1}=-(-1)^{n/2}c_{n-1}$ for odd order coefficients, etc.

\subsection{Sum rules for positive real functions}\label{sect:PRsumrules}
Based on \cite[Theorem 20.2 and 20.3]{Nedic+etal2019} we can now formulate the following definitions and the corresponding 
sum rules for PR functions.
A PR function $p$ is said to admit at $s=0$ an {\em odd} asymptotic expansion of odd order $M$ if for $M\geq -1$ there exist real numbers
$a_{-1},a_{1},\ldots,a_{M}$ such that $p$ can be written 
\begin{equation}\label{eq:PR_assympt1}
p(s)=a_{-1}s^{-1}+a_{1}s+\cdots+a_{M}s^{M}+o(s^{M}), \quad \textrm{as } s\toh 0.
\end{equation}
Similarly, a PR function $p$ is said to admit at $s=\infty$ an {\em odd} asymptotic expansion of odd order $M$ if for $M\geq -1$ there exist real numbers
$b_1,b_{-1},\ldots,b_{-M}$ such that $p$ can be written 
\begin{equation}\label{eq:PR_assympt2}
p(s)=b_1s+b_{-1}s^{-1}+\cdots+b_{-M}s^{-M}+o(s^{-M}), \quad \textrm{as } s\toh\infty.
\end{equation}
It can be shown that every PR function has an odd asymptotic expansion both at $s=0$ and at $s=\infty$ of order $-1$, 
and we have $a_{-1}=\lim_{s\toh 0}sp(s)=\beta(\{0\})\geq 0$ and $b_1=\lim_{s\toh\infty}s^{-1}p(s)\geq 0$.

For a positive real function to admit at $s=0$ an {\em odd} asymptotic expansion of odd order $M$ where $M\geq 1$,
it is both necessary and sufficient that the following sum rules (moments of the measure) hold
\begin{equation}\label{eq:Herglotz_SR1PR}
\int_{\R\setminus\{0\}}\frac{\mrm{d}\beta(\xi)}{\xi^{n}}
=\left\{\begin{array}{ll}
a_1-b_1 & n=2 \vspace{0.2cm} \\
-(-1)^{n/2}a_{n-1} & n=4,6,\ldots, M+1.
\end{array}
\right.
\end{equation}
As a consequence, we see also that $a_1\geq b_1$.
Similarly, for a positive real function to admit at $s=\infty$ an {\em odd} asymptotic expansion of odd order $M$ where $M\geq 1$,
it is both necessary and sufficient that the following sum rules (moments of the measure) hold
\begin{equation}\label{eq:Herglotz_SR2PR}
\int_{\R\setminus\{0\}}\xi^n\mrm{d} \beta(\xi)
=\left\{\begin{array}{ll}
b_{-1}-a_{-1} & n=0 \vspace{0.2cm} \\
(-1)^{n/2}b_{-n-1} & n=2,4,\ldots, M-1.
\end{array}
\right.
\end{equation}
As a consequence, we see also that $b_{-1}\geq a_{-1}$.
In \eqref{eq:Herglotz_SR1PR} and \eqref{eq:Herglotz_SR2PR} it is also possible to make the following identification
\begin{multline}
\lim_{\varepsilon\rightarrow 0+}\lim_{\sigma\rightarrow 0+}\frac{1}{\pi}\int_{\varepsilon<|\xi|<1/\varepsilon}\xi^{\pm n}\Re\{p(\sigma-\iu \xi )\}\mrm{d} \xi \\
=\int_{\R\setminus\{0\}}\xi^{\pm n}\mrm{d}\beta(\xi)
\end{multline}
as in \eqref{eq:betaprime}, \cf also \eg \cite[Eq.~(20.10)]{Nedic+etal2019} and \cite[Theorem 3.2.1]{Akhiezer1965}.
It is important to notice here that a possible point mass at $\xi=0$ is not included in the integrals expressed in \eqref{eq:Herglotz_SR1PR} and \eqref{eq:Herglotz_SR2PR}.

It is finally noted that the sum rules expressed in \cite[Theorem 20.2 and 20.3]{Nedic+etal2019} are given for general Herglotz functions without
any assumptions about symmetry. It is also noticed that these theorems require that the corresponding asymptotic expansion coefficients are {\em real valued} up to the required order.
The even ordered coefficients are purely imaginary for a symmetric Herglotz function, and hence follows the requirement of having an odd ordered asymptotic
expansion for symmetric Herglotz functions as well as for PR functions up to the required order, as in \eqref{eq:PR_assympt1} and \eqref{eq:PR_assympt2}.



\end{document}